\documentstyle[psfig,epsf,12pt]{article}
\newcommand{\nin}{\noindent}

\newcommand{\be}{\begin{equation}}
\newcommand{\ee}{\end{equation}}
\newcommand{\bea}{\begin{eqnarray}}
\newcommand{\eea}{\end{eqnarray}}

\begin{document}
\begin{center}
\vspace{15mm}
{\bf DYNAMICAL TUNNELING IN OPTICAL CAVITIES}
\vspace{5mm}

{Gregor Hackenbroich\\}
{\footnotesize{\it Yale University, Department of Applied Physics, 15 Prospect
Street, New Haven CT 06520, USA}}\\[2ex]
{Jens U.\ N\"ockel\\}
{\footnotesize{\it Max-Planck-Institut f{\"u}r Physik komplexer Systeme, 
Bayreuther Str.\ 40, D-01187 Dresden, Germany}}
\vspace{10mm}

{ABSTRACT}\\
\parbox[t]{30pc}{
\nin
{\footnotesize
The lifetime of whispering gallery modes in a dielectric cavity with a
metallic inclusion is shown to fluctuate by orders of magnitude when
size and location of the inclusion are varied. We ascribe these
fluctuations to tunneling transitions between resonances quantized in
different regions of {\em phase space}. This interpretation is
confirmed by a comparison of the classical phase space structure with
the Husimi distribution of the resonant modes. A model Hamiltonian is
introduced that describes the phenomenon and shows that it can be
expected in a more general class of systems.
}}
{{\em Published in} Europhys.~Lett.~{\bf 39}, 371 (1997)}
\end{center}

\vspace{5mm}
In recent years a lot of experimental effort
\cite{Yamamoto} has been devoted to dielectric microresonators. In
these resonators long-lived whispering gallery (WG) modes are created
by total internal reflection of light circulating just inside the
surface of the dielectric. WG modes can have $Q$ values exceeding
\cite{Gorodetsky} $10^9$ and they are of great interest for
applications such as microlasers and optical interconnects. For
optical cavities that are deformed from rotationally symmetric shape
it was demonstrated \cite{Noeckel} that the emission pattern of WG
modes can be understood in a ray-optics picture. The ray dynamics
typically exhibits both stable and chaotic trajectories, and it was
argued that the presence of chaos should lead to a broadening of the
WG modes with increasing deformation.
      
In this paper we study effects in dielectric cavities that are beyond
a ray-optics model. We show that tunneling transitions between
classically disconnected regions in phase space can lead to
fluctuations in the lifetime of WG modes by several orders of
magnitude. This is demonstrated numerically by solving the wave
equation for a dielectric that has a permeable coating and the shape
of the annular billiard. The fluctuations arise from
avoided crossings of WG modes with much broader resonances of the
dielectric.  The coating serves to make this effect more pronounced by 
raising all lifetimes above those obtained from total internal 
reflection alone. 
The Husimi projections of the relevant modes reveal that
the broad resonances and the WG modes are localized in chaotic and
regular regions of phase space, respectively. This motivates us to
model the dielectric in terms of a tunneling Hamiltonian where states
quantized in different regions of phase space are connected by small
tunneling matrix elements. We derive the quantum scattering matrix
associated with the tunneling Hamiltonian and obtain the widths of the
quasibound states. In agreement with our numerical observation we find
strong fluctuations in the lifetime of WG modes close to avoided
crossings with modes localized in a different region of phase space.

The idea that disconneted regions in classical phase space can give
rise to quantum tunneling has previously been introduced by Davis and
Heller \cite{DH}, who dubbed this notion dynamical tunneling.
Dynamical tunneling has been observed
\cite{Bohigas1,Bohigas2,Leyvraz,Eyal} in the spectrum of
{\em closed} quantum systems whose classical counterparts exhibit both
regular and chaotic motion. In such systems dynamical tunneling leads
to statistical fluctuations in the splittings of nearly degenerate
energy levels. While the resulting level splittings are typically so
small that their experimental detection seems unlikely, we believe
that the tunneling effects giving rise to fluctuations in resonance
lifetimes of {\em open} systems are amenable to experimental
verification. 

We numerically study a dielectric cavity with the shape of the
annular billiard \cite{Bohigas2,Eyal}. The system [Fig.~1 a)] consists
of a cylindrical dielectric with radius $R$ and index of refraction
$n>1$ that is suspended in air. Embedded in this cylinder and
off-center by an amount $\delta$ is a totally reflecting (metallic)
rod of radius $a<R-\delta$. The surface of the dielectric is covered
by a penetrable metallic film. The dielectric function of the system
is written as $\epsilon = \epsilon^{\rm bulk}+\epsilon^{\rm film}$
where $\epsilon^{\rm bulk}=n^2$ inside the dielectric and
$\epsilon^{\rm bulk}=1$ outside. The metallic film is taken as a
$\delta$-function layer, $\epsilon^{\rm film}=-\eta\,n^2R\,\delta(r-R)$
with $\eta>0$, corresponding to a purely imaginary refractive
index. This neglects any phase shifts in the film (caused by a real
part of the index), as well as the frequency-dependence of the
absorption.  Absorption by the inclusion is neglected because we
expect the dominant loss to be caused by the outer interface.

We solve for the quasibound states of the annular billiard
using wave function matching. The electric field $\Psi$ when
polarized parallel to the cylinder axis satisfies the scalar wave
equation
\begin{equation}  
\label{wave}  
\nabla^2 \Psi+\epsilon({\bf r}) k^2 \Psi = 0 , 
\end{equation}
and is continuous at $r=R$. The $\delta$-function in $\epsilon$
creates a jump in the derivative, $\partial \Psi/ \partial r |_R =
\eta\,n^2\,k^2 R\,\Psi(R)$. We search for a solution {\em with no
incoming waves},
\begin{eqnarray}
\label{ein} && \Psi_< \!=\!\! \sum_\mu A_\mu \Psi_\mu^+, 
\quad \Psi_>=\sum_\mu B_\mu H_\mu^{(1)} (kr) \cos \mu \phi, \\
&& \label{eout} \Psi^+ = H_\mu^{(2)} (nkr) \cos \mu \phi  \! + \!  
\sum_\nu  S_{\nu \mu}^{(I)} H_\nu^{(1)}\! (nkr) \cos \nu \phi,
\end{eqnarray} 
where $\Psi_<$ and $\Psi_>$ denote the electric field for $r<R$ and
$r>R$, respectively. We use polar coordinates $r$, $\phi$ with origin
in the center of the dielectric, and $H_\mu^{(1,2)}$ are the Hankel
functions of first and second kind. The scattering
matrix $S^{(I)}$ of the inner annulus is known explicitly
\cite{Eyal}. Matching $\Psi$ and $\partial \Psi/ \partial r$ at the
surface $r=R$, one obtains a set of simultaneous equations for the
Fourier coefficients $A_\mu, B_\mu$, which can only be solved at
discrete complex values of the wavenumber $k$. To each solution
we assign the radial and angular momentum quantum numbers ($u$, $m$)
of the state in the concentric billiard ($\delta = 0$) from which it
evolved.
\begin{figure}[bt]
\vspace*{2cm}
\hbox{\hspace*{2cm}\psfig{figure=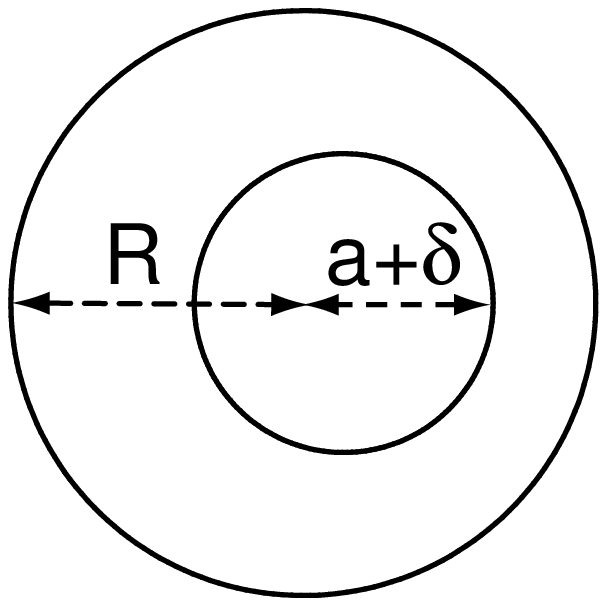,height=2cm}}
\vspace*{-4cm}
\hbox{\hspace*{4cm}
\psfig{figure=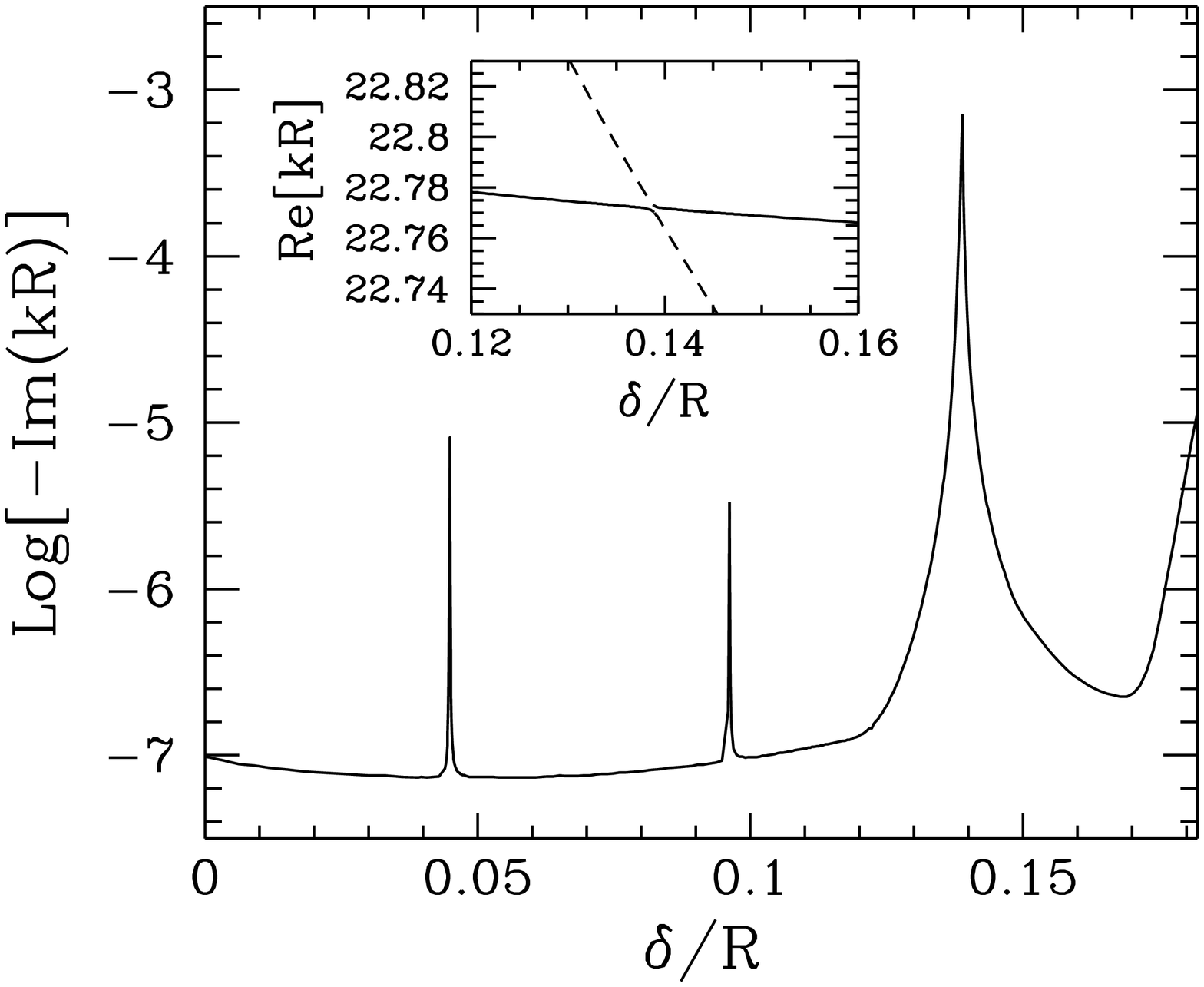,height=8cm,angle=270}}
\vspace*{-0.5cm}

\noindent\footnotesize
a) Cross seection of cylindrical resonator with off-center 
metallic inclusion (inner disk). b)
Width of the WG mode (2,30) vs.\ eccentricity $\delta/R$. 
Inset: Positions of WG mode (2,30) (solid line) and the closest broad resonance
(dashed). 
\normalsize
\end{figure}

The {\em closed} annular billiard ($\eta \to \infty$) exhibits mixed
classical dynamics \cite{Bohigas2}. Let $\chi$ be the {\em angle of 
incidence} measured from the surface normal. Then WG
trajectories are those which never hit the inner cylinder, i.e.
their impact parameter is $R\,|\sin\chi| > a+\delta$.
Their motion is regular, with a {\em conserved} $\chi$.
The phase space for $R\,|\sin\chi|<a+\delta$ contains both
regular and chaotic regions if $\delta>0$. For chaotic rays, $\chi$ 
fluctuates between reflections and can become very small.
In the {\em open} system, this means that 
modes supported by the chaotic domain have widths that are 
several orders of magnitude broader than the widths of WG modes,
because the outer coating is far more penetrable close to normal 
incidence than at the grazing angles allowed for WG rays.

Fig.~1 b) shows the width of the even WG mode (2,30) corresponding to
$\sin \chi = 0.66$ (at $\delta = 0$) as a function of $\delta/R$. We
chose $n=2$, $\eta = 1$ and kept $a+\delta=0.65$ fixed. This ensures
that WG modes with $|\sin \chi| >(a+\delta)/R$ remain classically
undisturbed. The width is seen to fluctuate by orders of magnitude.
This is caused by avoided crossings with broader resonances, as we
show in the inset for the large peak at $\delta/R \approx 0.139$
\cite{width}. Here, the dashed line corresponds to a broad resonance
of width $-{\rm Im} (kR) \approx 10^{-3}$.

\begin{figure}[hbt]
\hbox{\hspace*{4cm}{\sf a)}\hspace*{6.55cm} {\sf b)}}
\hbox{\hspace*{1cm}\epsfbox[0 0 200 135]{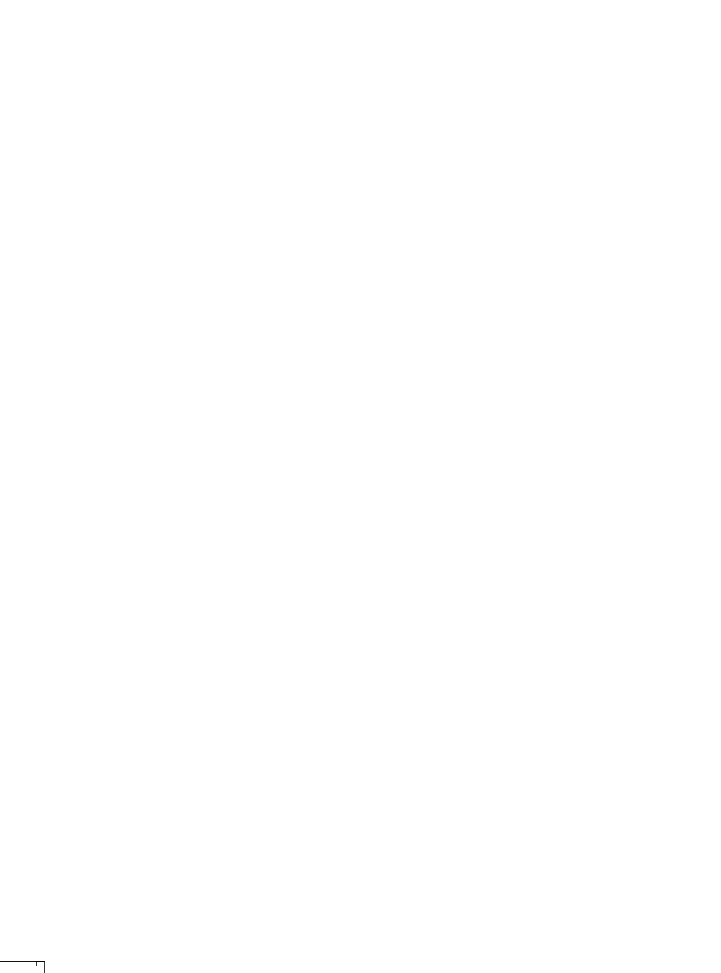}
\epsfbox[0 0 200 135]{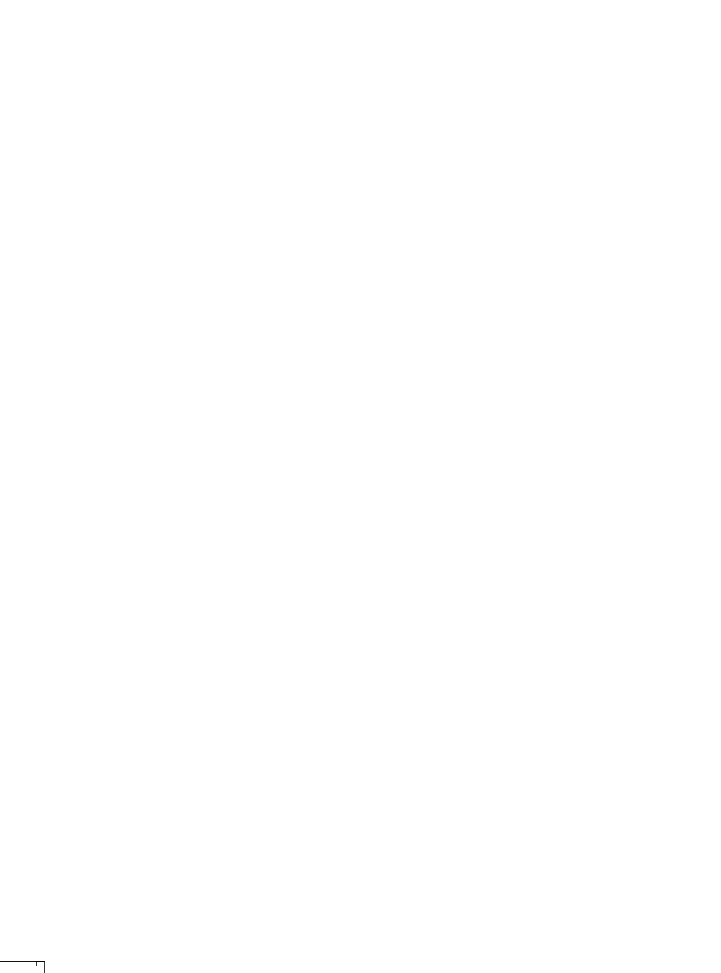}}
\hbox{\hspace*{7.5cm}{\sf c)}}
\hbox{\hspace*{4.5cm}\epsfbox[0 0 60 135]{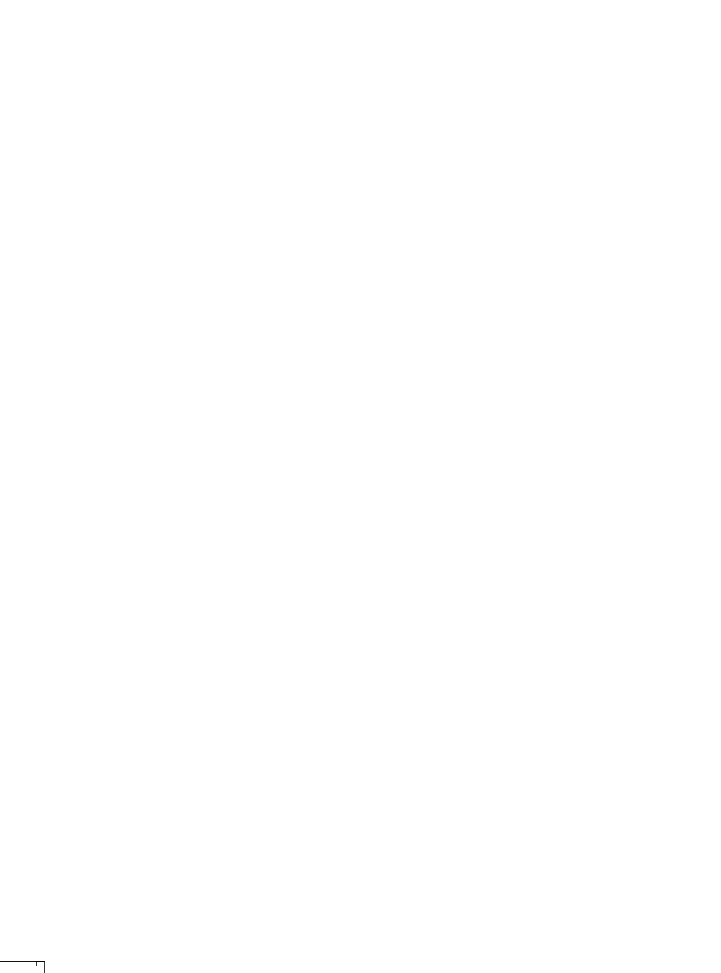}}
\vspace*{5mm}
\noindent\footnotesize
The projection $\rho_{SOS}$ of the Husimi function onto
the Poincar\'e surface of section, superimposed with classical
trajectories to indicate the extend of regular and chaotic
regions. The uncertainty in $\phi$ is 0.5 radians; $\delta / R =
0.135$ in a) and b), with a) showing the phase space density of the WG
mode and b) that of the broad resonance. In c), $\rho_{SOS}$ is shown
for the WG state at $\delta /R = 0.139$.
\normalsize
\end{figure}
We now want to associate these resonances with different regions of
the classical phase space spanned by $r$, $\phi$ and their conjugate
momenta $(p_r,p_\phi)$. This is achieved using the Husimi function,
i.e.\ the overlap of $\Psi_<$ with a minimum-uncertainty wavepacket
$\Phi$ centered at $\bar{r},\bar{\phi},\bar{p}_r,\bar{p}_\phi$. The
resulting phase space density $\rho_H(\bar{r},\bar{\phi},
\bar{p}_r,\bar{p}_\phi)$ characterizes $\Psi_<$. To obtain a classical
phase-space portrait, we employ a Poincar\'e surface of section with
$\phi$ and $\sin \chi$ as coordinates. By semiclassical arguments,
$\sin \chi$ can be directly related to $p_\phi$. One can then project
$\rho_H$ onto the surface of section by integrating out $\bar{p}_r$,
choosing the spread of $\Phi$ around $\bar{r}$ to be zero and setting
$\bar{r}=R$.

The resulting density distribution $\rho_{SOS}$ is shown in Fig.~2 for
the two resonances singled out in the inset to Fig.~1 b). Away from the
avoided crossing, the WG mode is localized in the regular regime at
$\sin \chi \approx 0.66$, whereas the broad resonance is supported in
the chaotic domain. However, close to the anticrossing at $\delta/R
\approx 0.139$, $\rho_{SOS}$ [Fig.~2 c)] reflects a strong coupling
between the regular and chaotic parts of phase space due to
interference of the two resonances. In contrast to Figs.~2 a) and b),
this is a {\em classically forbidden} phase space distribution made
possible only by dynamical tunneling. It causes the lifetime of the WG
mode to approach that of the chaotic state. We find fluctuations
similar to Fig.~1 b) for all WG modes with $|\sin \chi | >
(a+\delta)/R$. 

Our analytical approach is motivated by the observation that the modes
of the cavity have overlap with distinct regions of the classical
phase space. Exploiting the analogy between the two-dimensional wave
equation and the two-dimensional Schr\"odinger equation we model the
system by the following Hamiltonian
\begin{equation}    
 H = H_0 + H_T.
\label{Hamiltonian}   
\end{equation}     
Here, $H_0$ describes states quantized in two classically disconnected
regions in phase space as well as the channel region when these
subsystems are {\em totally disconnected} from each other. The
couplings between the subregions are described by $H_T$. Written in a
basis where the various subsystems are diagonal, $H_0$ has the form
\begin{equation}    
H_0 = \sum_i {\cal E}_i q_i^\dagger q_i^{} +
\sum_\mu E_\mu  c_\mu^\dagger c_\mu^{}  + \sum_a \int dE E
d_{aE}^\dagger d_{aE}^{} .   
\label{H0}  
\end{equation}   
Here, the creation (annihilation) operators $q_i^\dagger$ and
$c_\mu^\dagger$ ($q_i$ and $c_\mu$) arise from the quantization of the
two respective regions of phase space. The corresponding quantum
states will be referred to as regular and chaotic states below.
A generalization of $H_0$ to include more regions in phase space is
straightforward. There is a continuum of channel states with
corresponding operators $d_{aE}^\dagger$, $d_{aE}^{}$, where
$a=1,2,\ldots,M$ denotes the channels. The tunneling Hamiltonian
$H_T$ is given by
\begin{eqnarray}   
H_T  =  &&\left( \sum_{ai} \int \! dE U_{ai}(E) d_{aE}^\dagger q_i
+H.c. \! \right)  +  \nonumber \\  
&& + \left( \! \sum_{i\mu} V_{i\mu} q_i^\dagger
c_\mu^{} + H.c. \right) +\left(\sum_{a\mu}
\int dE  W_{a\mu}(E) d_{aE}^\dagger  c_\mu^{}+H.c. \right).    
\end{eqnarray}   
It couples both the regular states and the chaotic states directly to
the continuum. Moreover, there is a coupling between the regular and
the chaotic states via the matrix elements $V_{i\mu}$.

The Hamiltonian (\ref{Hamiltonian}) previously has been
investigated in various limiting cases. The case $V_{i\mu}=U_{ai}=0$
together with the assumption that the matrix elements $W_{a\mu}$ be
random variables defines a model for chaotic scattering which has been
the subject of intensive research in the past decade
\cite{Verbaarschot,Fyodorov}. The case of vanishing channel couplings,
$U_{ai}=W_{a\mu}=0$ has been studied in recent work [4-6] on the
splitting distribution in {\em closed} quantum systems.  Here, we
shall investigate the effects of tunneling transitions between the
regular and the chaotic region ($V_{i\mu} \neq 0$) when these regions
are weakly coupled to the channels. The channel couplings are assumed
so weak that the typical widths of both the regular and the chaotic
resonances are smaller than their respective energy spacing (the
regime of isolated resonances).
 
We calculate the scattering matrix $S_{ab}(E)=\delta_{ab}- 2\pi i
T_{ab}$ at energy $E$ from the Lippmann-Schwinger equation   
\begin{equation}    
T = H_T +H_T {1 \over E - H_0+i \eta}T  
\label{Lippmann}
\end{equation} 
for the transition operator T ($\eta$ is positive infinitesimal).
This equation is solved by the explicit summation of the associated
Born series \cite{Hackenbroich}.  Here it is sufficient to discuss the
solution in the case that the direct coupling between regular and
channel states is much weaker than the coupling arising from indirect
processes involving chaotic states.  One finds that the $S$ matrix has
the form $S_{ab}= \delta_{ab}-2\pi i T^{C}_{ab}-2\pi i T^{R}_{ab}$
where $T^{C}_{ab}$ comprises all scattering processes involving only
chaotic states while all remaining processes are included in
$T^{R}_{ab}$. The poles of $T^{R}_{ab}$ are located at complex
energies $E_i =\epsilon_i -(i/2)\Gamma_i$, where $\epsilon_i$ is the
position and $\Gamma_i$ the width of the i-th resonance. Assuming that
this resonance is isolated from all other resonances, the result is
\begin{equation}  
\label{Gam1}
\Gamma_i = 2 \pi \sum_{a} \left| \sum_\mu { W_{a\mu}^{} V_{i\mu}^*
\over \epsilon_i-E_\mu+i\pi \sum_b W_{b\mu}^* W_{b\mu}^{}}
\right|^2, 
\end{equation} 
where the matrix elements are evaluated at energy $\epsilon_i$. This
resembles the expression for the transition rate in second-order
perturbation theory, but in contrast to the perturbative expression
the denominator in Eq.~(\ref{Gam1}) has a finite imaginary part. 
In accordance with our
numerical results for the lifetime of WG modes, Eq.~(\ref{Gam1}) shows
a large increase of the width whenever a regular WG mode (with index
$i$) crosses a chaotic mode (with index $\mu$). In this way, 
multi-step tunneling processes involving chaotic states cause 
strong fluctuations in the resonance widths.

It is known \cite{Friedrich} that the interference of resonances may
strongly affect their positions and widths. To study this effect
very close to an avoided crossing we consider the nonhermitian
Hamiltonian  
\begin{equation}    
\label{inter} 
H = \left(
\begin{array}{cc}
{\cal E}_i & V \\
V^* & E_\mu
\end{array} \right) -i
\left(
\begin{array}{cc}
0 & 0 \\
0 & \bar{\Gamma} /2 
\end{array} \right)
\end{equation}
as a simple model of two resonances that are coupled to single open
channel. If the separation $|{\cal E}_i-E_\mu|$ is much larger than
$|V|$, there are two eigenstates with width close to $0$ and
$\bar{\Gamma}$, respectively. For small values of $|{\cal E}_i
-E_\mu|$, one finds an avoided crossing of the resonance positions
(the real parts of the eigenvalues). At the same time, both widths
approach the value $\bar{\Gamma} /2$. We expect the sharp resonance
typically to aquire a width of order $\bar{\Gamma}$ (where
$\bar{\Gamma}$ is the width of the broader resonance when both
resonances are separated) even in the general case of arbitrary number
of open channels. This is clearly reflected in our numerical data
presented in Fig.~1 b).

Further progress can be made assuming that the quantum states
localized in the chaotic regions of phase space obey random-matrix
behavior. In this case, the width (\ref{Gam1}) becomes a statistical
quantity and one can compute its probability distribution
$P(\Gamma_i)$ over a large number of anticrossings. Due to avoided
crossings, this distribution shows a power-law decay $P(\Gamma_i) \sim
\Gamma_i^{-3/2}$ in the regime $\Gamma_i \ll \bar{\Gamma}$, where
$\bar{\Gamma}$ denotes the average width of the chaotic resonances.
The derivation of the distribution $P(\Gamma_i)$ and a further
discussion is defered to a forthcomming publication \cite{HN}.

In summary, we have studied the lifetime and the phase space
distribution of WG modes in a dielectric cavity. Dynamical tunneling
leads to large fluctuations of the lifetimes as a function of the
asymmetry parameter, as we demonstrated numerically for the open
annular billiard. This system can be realized using dielectric
microcavities.

We thank E.\ Doron and S.\ Frischat for helpful discussions and for
communicating unpublished material. We acknowledge discussions with
A.\ D.\ Stone and H.\ A.\ Weidenm\"uller.  This work was supported the
Alexander von Humboldt Foundation and by NSF Grant No.\ DMR-9215065.


\begin{thebibliography}{99}
\bibitem{width} The FWHM of the resonance width is of the
same order as the width determined from the avoided crossing. In
Fig.~1 b) the resonance width of the broad peak appears to be larger than
the width of the avoided crossing due to the fact that we plot the
logarithm of the width.
\end{thebibliography}
\end{document}